\pdfoutput=1
\documentclass[11pt]{article}
\usepackage{acl}
\usepackage{times}
\usepackage{latexsym}
\usepackage[T1]{fontenc}
\usepackage[utf8]{inputenc}
\usepackage{url}
\usepackage{amsmath,amssymb,amsfonts}
\usepackage{graphicx}
\usepackage{textcomp}
\usepackage{xcolor}
\usepackage{times}
\usepackage{algorithm}
\usepackage{hyperref}
\usepackage{latexsym}
\usepackage{amsmath}
\usepackage{mathtools}
\usepackage{listings}
\usepackage{caption}
\usepackage{subcaption}
\usepackage{stmaryrd}
\usepackage{enumitem}
\usepackage{xspace}
\usepackage{float}
\usepackage{soul}
\usepackage{afterpage}
\usepackage{tikz}
\usepackage[]{footmisc}

\usepackage{microtype}
\usepackage{inconsolata}
\definecolor{jccolor}{rgb}{0.1,0.7,0.8}
\definecolor{vlcolor}{rgb}{0.9,0.1,0.1}
\definecolor{gcolor}{rgb}{0.7,0.3,0.7}
\definecolor{ccolor}{rgb}{0.3,0.3,0.7}
\definecolor{mrcolor}{RGB}{200,30,50}
\definecolor{mscolor}{RGB}{8, 102, 3}
\definecolor{bzcolor}{rgb}{0.05, 0.9, 0.2}
\definecolor{ntcolor}{RGB}{186, 85, 211}
\definecolor{encolor}{RGB}{219, 112, 147}
\definecolor{jwcolor}{RGB}{220, 20, 60}
\definecolor{ascolor}{RGB}{218, 165, 32}
\definecolor{red}{RGB}{200, 0, 0}
\definecolor{npcolor}{RGB}{255, 0, 255}

\newcommand{\eg}{\emph{e.g.\@}\xspace}
\newcommand{\ie}{\emph{i.e.\@}\xspace}

\usepackage{algorithm}
\usepackage[noend]{algpseudocode}
\usepackage{listings}
\algrenewcommand\algorithmicensure{\textbf{Output:}}
\algrenewcommand\algorithmicrequire{\textbf{Input:}}

\newcommand{\pass}[1]{pass@$#1$}
\newcommand{\partialpass}[2]{pass@$#1$($#2$\%)}

\newcommand{\query}{\ensuremath{Q}}
\newcommand{\valset}{\ensuremath{\textsc{Validate}}}
\newcommand{\prompt}{\ensuremath{P}}
\newcommand{\sol}{\ensuremath{c}}
\newcommand{\solset}{\ensuremath{C}}
\newcommand{\kvar}{\ensuremath{k}}
\newcommand{\maxcomp}{\kvar_{max}}
\newcommand{\budget}{\ensuremath{B}}

\newcommand{\cache}{\ensuremath{O}}

\newcommand{\num}{\ensuremath{n}}

\newcommand{\llm}{\ensuremath{\textsc{Llm}}}
\newcommand{\codex}{\tname{Codex}}
\newcommand{\incoder}{\tname{InCoder}}
\newcommand{\codellama}{\tname{CodeLlama}}

\newcommand{\gptthree}{\tname{Gpt-3}}
\newcommand{\gptfour}{\tname{Gpt-4}}

\newcommand{\spider}{\tname{Spider}}

\newcommand{\suchthat}{{\mbox{s.t.\ }}}

\newcommand{\set}[2]{\left\lbrace\,#1 \suchthat #2\,\right\rbrace}

\newcommand{\dataframe}{\ensuremath{T}}
\newcommand{\rows}{\ensuremath{R}}

\newcommand{\clustermap}{\ensuremath{M}}

\newcommand{\cluster}{\ensuremath{\textsc{Cluster}}}
\newcommand{\select}{\ensuremath{\textsc{Select}}}

\newcommand{\emphbf}[1]{\emph{\textbf{#1}\xspace}}
\newcommand{\mypara}[1]{\smallskip\noindent\emphbf{#1.}\xspace}
\newcommand{\myparabf}[1]{\smallskip\noindent\textbf{#1}\xspace}

\newcommand{\scode}[1]{{\texttt{#1}}}
\newcommand{\tcode}[1]{{\texttt{\small #1}}}

\newcommand{\tname}[1]{\textsc{#1}\xspace}
\newcommand{\jigsaw}{\tname{Jigsaw}}
\newcommand{\palm}{\tname{PaLM}}
\newcommand{\alphacode}{\tname{AlphaCode}}
\newcommand{\humaneval}{\tname{HumanEval}}
\newcommand{\apps}{\tname{APPS}}
\newcommand{\cert}{\tname{Cert}}
\newcommand{\sof}{\tname{SofSet}}

\newcommand{\ind}{\tname{ind}}
\newcommand{\dep}{\tname{dep}}
\newcommand{\ext}{\tname{ext}}

\iffalse 
\newcommand{\nbc}[3]{}
\else

\newcommand{\nbc}[3]{
	{\colorbox{#3}{\bfseries\sffamily\scriptsize\textcolor{white}{#1}}}
	{\textcolor{#3}{\sf\small \textit{#2}}}
}
\fi

\title{Solving Data-centric Tasks using Large Language Models}
\author{
Shraddha Barke\textsuperscript{*}, 
Christian Poelitz\textsuperscript{†}, 
Carina Suzana Negreanu\textsuperscript{†}, 
Benjamin Zorn\textsuperscript{†}, 
José Cambronero\textsuperscript{†},\\
Andrew D. Gordon\textsuperscript{†}, 
Vu Le\textsuperscript{†}, 
Elnaz Nouri\textsuperscript{†}, 
Nadia Polikarpova\textsuperscript{†}, 
Advait Sarkar\textsuperscript{†}, 
Brian Slininger\textsuperscript{†}, 
Neil Toronto\textsuperscript{†}, \\
Jack Williams\textsuperscript{†}
{\small \textsuperscript{*}\texttt{\{sbarke, npolikarpova\}@ucsd.edu}} \\
{\small \textsuperscript{†}\texttt{\{cpoelitz, cnegreanu, Ben.Zorn, jcambronero, adg, levu, Elnaz.Nouri, advait, brslinin, neil.toronto, jack.williams\}@microsoft.com}}
}

\begin{document}
\maketitle
\begin{abstract}
Large language models are rapidly replacing help forums like StackOverflow, and are especially helpful to non-professional programmers and end users.
These users are often interested in \emph{data-centric tasks}, like spreadsheet manipulation and data wrangling, which are hard to solve if the intent is only communicated using a natural-language description, without including data.
But how do we decide how much data and which data to include in the prompt?

This paper makes two contributions towards answering this question. 
First, we create a dataset of real-world NL-to-code tasks manipulating tabular data,
mined from StackOverflow posts. 
Second, we introduce a novel \emph{cluster-then-select} prompting technique, which adds the most representative rows from the input data to the LLM prompt.
Our experiments show that LLM performance is indeed sensitive to the amount of data passed in the prompt,
and that for tasks with a lot of syntactic variation in the input table,
our cluster-then-select technique outperforms a random selection baseline.
\end{abstract}
\section{Introduction}\label{sec:intro}
\begin{figure*}[t]
\centering
\footnotesize
\includegraphics[width=\textwidth, height=0.55\textwidth]{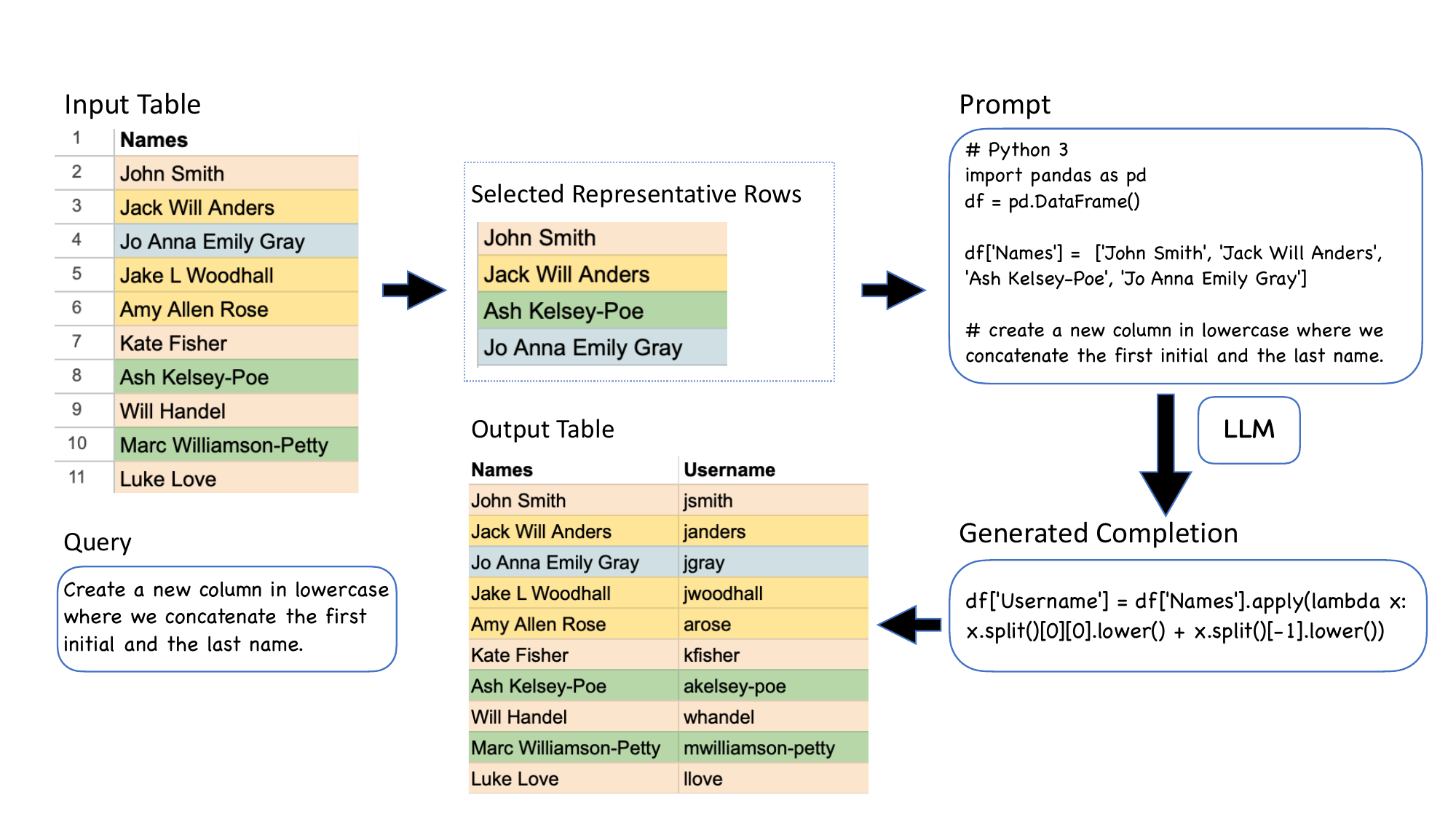}
\caption{An overview of our \emph{cluster-then-select} prompting technique. The input is a data table and natural language query.
The rows in the data table are first clustered based on their syntactic structure (in this case the name format). We depict different clusters using distinct colors. The most representative rows are then selected from each cluster to create a prompt to pass to the model. Finally, the generated completion is used to create an output column.}
\label{fig:example}
\end{figure*}

Code-generating large language models (LLMs) promise to empower end users interested in \emph{data-centric tasks},
ranging from string manipulations in spreadsheets to data cleaning and analysis in computational notebooks.
For example, consider the following task on tabular data:
given a column with \emph{full names},
generate a new column with \emph{user names},
by combining the first initial and last name, in lowercase.
This task can be solved by a Pandas program that:
1) splits the full name into a list of strings, 2) extracts the first and last string from the list, 3) converts both to lowercase and joins the first letter of one string to the other as shown in \autoref{fig:example}.
The challenge in generating this program
is that input data rows have varied formats, \eg
most rows only have two names ("John Smith"), 
but some have multiple middle names ("Jake L Woodhall", "Jo Anna Emily Gray").
If an LLM prompt does not include any data
or only includes rows with two names,
the LLM is more likely to generate a program that does not generalize
(\eg one that extracts the last name as the \emph{second} element of the list instead of \emph{last}).

In this paper, we focus on solving such tasks that involve multi-step computations on the input columns to generate additional columns.
Towards this goal, we mine StackOverflow to construct a new dataset, dubbed \sof,
of data-centric tasks, equipped with a natural-language query
and a small input table.
%
%
Using this dataset, we conduct experiments on generating Pandas programs using \gptfour and an open-source alternative \codellama,
with the goal of analysizing LLMs' sensitivity to the amount of input data provided in the prompt.

Unlike input tables in StackOverflow posts,
real-world data tables are often large,
hence sending the entire table to the LLM is likely 
impractical, expensive, or detrimental to performance.
%
\emph{How do we best convey the structure of a large input table to the LLM?}
To address this question, we propose a \emph{cluster-then-select} prompting technique that clusters input rows based on their syntactic structure and then selects representative rows from each cluster;
\eg in our \autoref{fig:example} example,
the technique would include a row for each format of middle names.
To evaluate this technique, we perform experiments on \sof
augmented with larger input tables extracted from Kaggle.
%
%

In summary, this paper contributes:
\begin{itemize}
\item a real-world dataset of complex tasks for evaluating data-centric code generation;

\item a \emph{cluster-then-select} technique for selecting rows to prompt with, from large input tables;

\item an analysis that shows LLMs are sensitive to the data quantity, choice and position of rows.
\end{itemize}
\section{Related Work}
\myparabf{Large language models for tabular data}
Code-generating LLMs like \codex \cite{https://doi.org/10.48550/arxiv.2107.03374}, \incoder \cite{fried2022incoder} and \palm \cite{PaLM} have been fine-tuned for code-specific tasks and adapted for data-centric domains like SQL \cite{CodexDB, TtoSQL}.
\cite{li2020deep} explore the ability of models like BERT to perform entity matching on tabular data.
\cite{narayan2022can} use \gptthree for data cleaning, error detection and entity matching tasks. 
\cite{pmlr-v206-hegselmann23a} focus on tabular classification tasks and explore parameter-efficient LLM tuning.

\myparabf{Prompting for data-centric tasks}
Prompting LLMs has been quite effective in practice across domains \cite{reynolds2021prompt, wang2022no, liu2023pre}. 
In this paper, we ask the question: \textit{how does data context impact code generation for data-centric tasks?}
Previous works have explored prompting with data: \cite{jain2022jigsaw} provide both input and expected output tables (which might not be available in a realistic setting).
\cite{gemmell2023generate} prompt with transformed tables after filtering out rows that are not relevant, for their question-answering tasks.
\cite{ye2023large} decompose a huge table into a smaller one, and convert a question into simpler sub-questions for tabular question-answering tasks.
\cite{pmlr-v206-hegselmann23a} serialize data tables into a textual representation for tabular classification tasks.
These works prompt LLMs for data analysis, classification and wrangling tasks (in-place data transformations) whereas we focus on multi-step data manipulation.
We propose a new \emph{cluster-then-select} prompting technique that clusters the input data and adds representative rows to the prompt.
\cite{yin2022natural} focus on data-centric tasks in
computational notebooks.
\section{The \textsc{SofSet} Dataset}
We collect a new dataset fashioned from real-world data-centric tasks from StackOverflow (\sof). 
We sample tasks deterministically from the highest rated posts with the tag "ExcelFormulas" in StackOverflow (as of March 2022). 
These tasks are representative of real problems spreadsheet users face frequently since they correspond to the highest-rated posts. 
We manually check that the posts are genuine tasks and also remove post identifiers for anonymization. 
This gives us a total of 201 tasks.
\subsection{Dataset Annotation}
Each datapoint in our dataset is annotated with a concise textual query, a data input (column-major-flat table), an expected correct output (extra columns), a pandas solution and metadata.
We manually write the textual queries, 
summarising the original verbose StackOverflow question. 
Each query is annotated and verified by at least three internal annotators. 
%
For the data input, we use the table from the original StackOverflow post (if present),
and manually add extra rows and corner cases until we have at least 10 rows.
%
%
Since the natural language query and tabular data are not verbatim copies from StackOverflow and we have a different target language for generation (Pandas instead of Excel Formulas), the evaluation data should not be present in the training data.
We choose Pandas as the target language since LLMs are especially good at generating Python but our methods and dataset are programming-language agnostic.
%
\subsection{Dataset Properties}
\emph{What makes our dataset different from existing ones?}
First, our dataset consists of complex data-centric tasks with multiple input columns. 
Python datasets like \apps \cite{Hendrycks2021MeasuringCC} and \humaneval\cite{https://doi.org/10.48550/arxiv.2107.03374} are not data-centric. 
Second, our dataset is larger than existing data-centric datasets: \jigsaw \cite{jain2022jigsaw} and \cert \cite{Zan2022CERTCP}.
\jigsaw has 79 unique tasks (median of 7 data rows) and \cert has 100 unique tasks (median of 3 rows).
Our dataset has 201 unique tasks, with a median of 10 rows.
%
The \spider dataset~\cite{DBLP:conf/emnlp/YuZYYWLMLYRZR18} is a text-to-SQL dataset which focuses on relational query tasks whereas we focus on fine-grained data wrangling and manipulation tasks.
Finally, we propose a \emph{taxonomy} of data-centric tasks,
classifying them into data-independent (\ind), data-dependent (\dep), and external-dependent (\ext),
based on the data required to produce a solution.

\myparabf{Data-independent tasks}
These tasks can be solved using the query alone without any data access. 
An example is the query 
"create a new column that includes only the first 5 characters from Filename".

\myparabf{Data-dependent tasks}
These tasks cannot be solved using the query alone: 
the model needs access to the input table. 
For example, the query "create a new column with the number of days between the two date columns" requires data access to identify the correct column names and date format, both absent from the query.

\myparabf{External-dependent tasks}
These tasks can only be solved with external world knowledge in addition to data access. The query "create a new column that counts how many US holidays are between the dates in Start Date and End Date", requires the model to know about US holidays.

Following this taxonomy, \sof consists of 126 \ind tasks, 44 \dep tasks and 31 \ext tasks.
These tasks span diverse domains including string manipulation, date and time, math, address, and complex conditionals among others.
%

\subsection{Cluster-then-select prompting technique}

%
To solve tasks on large tables, we propose a \emph{cluster-then-select} technique which prompts the model with a representative sample of the input data.
In order to capture the syntactic variation in the input data, we rely on an existing tool~\cite{padhi2018flashprofile}, which takes as input a set of strings and synthesizes a small set of regular expressions (regexes), such that each input string matches one of the regexes.
In our example in \autoref{fig:example}, 
it would synthesize separate regexes for rows with zero, one and two middle names and hyphenated last name.
Names like "John Smith" would belong to the zero middle name cluster and "Jack Will Andres" and "Jo Anna Emily Gray" belong to the clusters with one and two middle names resp. 
Also, the name "Ash Kelsey-Poe" would belong to the cluster with hyphenated last names.
These regexes are then used to cluster the input strings, and we select some number of rows from each cluster.
%
In \autoref{fig:example}, we pick one row from each of the four distinct clusters (depicted with different colors).

If the input table only has one column, selecting $n$ representative rows based on the clustering results is trivial: simply pick one row each from the top-$n$ most populous clusters.
In cases where the input contains more than one column, they may be clustered differently. 
We then select $n$ rows that together cover as many strings as possible across all the columns.
We frame this as a \emph{weighted maximal coverage problem}~\cite{maxcoverage}, which can be solved approximately in a greedy manner.
In each iteration, the algorithm selects the row whose elements maximize cluster coverage.

\myparabf{Kaggle-augmented dataset}
In order to evaluate our \emph{cluster-then-select} technique on larger datasets, we expand the 44 data-dependent tasks by adding more rows from open-source Kaggle datasets~\cite{kaggle}, bringing the total to 1000 rows.
We first identify the data domains in the original \sof rows (such as names, numbers, address, date, time etc) and then source comparable open-source Kaggle datasets of the same domain.
We further post-process the Kaggle data to maintain the original rows format and ensure that the augmented data is coherent.
This introduces greater variation in the original data which increases the number of data clusters.
62\% of our \dep tasks have at least two clusters and we have tasks with up to ten clusters.
%
Since the Kaggle data is post-processed and is not tied to the task query in any way, it is unlikely to bias the LLM evaluation by being part of the training data.
This larger dataset allows for a thorough evaluation, better mirroring real-world conditions.
\begin{figure}[t]
\centering
\includegraphics[width=0.5\textwidth, height=0.4\textwidth]{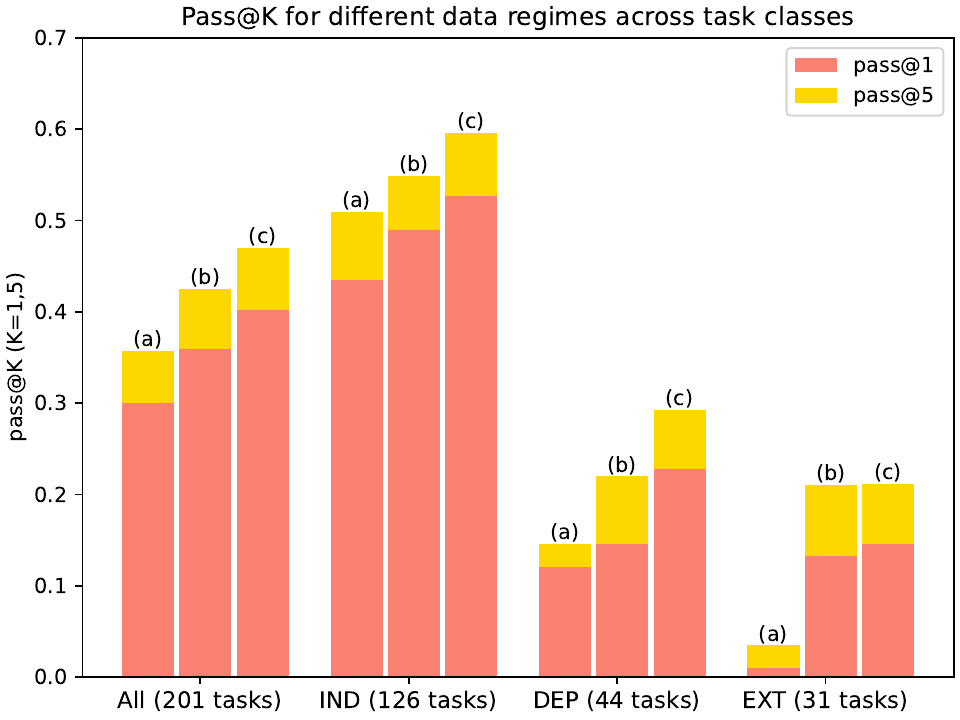} 
\caption{\pass{k} with (a) no-data, (b) first-row, and (c) ten-rows passed to the model. The leftmost group of bars represent pass@k with all classes followed by separate pass@k for \ind, \dep and \ext tasks.}
\label{fig:sof-data}
\end{figure}
%
\section{Evaluation of data-centric tasks}
We perform an analysis of the role of data on model performance in data-centric tasks.
We first use the original \sof dataset to examine three data regimes with increasing amounts of data: (a) no-data (b) first-row and (c) ten-rows and also the taxonomy of task classes of increasing difficulty in terms of data required: \ind, \dep and \ext.
We then use Kaggle-augmented \dep tasks to compare our \emph{cluster-then-select} technique (which selects representative rows from the top-n most dense clusters) against a random baseline (which selects random rows from the input table).
For each data setting, we construct a prompt which contains the task query and selected rows as a pandas dataframe to generate code from \gptfour as shown in \autoref{fig:example}.
Correctness is reported based on whether the code produces the expected output in terms of \pass{k}, the probability that at least one of k samples of generated code produces the correct output~\cite{https://doi.org/10.48550/arxiv.2107.03374}.
We report all results using \gptfour with a temperature of 0.5 and the generated completions are evaluated on all rows in the input table. 
The \sof dataset, all the evaluation results and our prototype tool can be found online.\footnote{\url{https://github.com/microsoft/CodeXData}}
\begin{figure}[t]
\centering
\includegraphics[width=0.5\textwidth, height=0.4\textwidth]{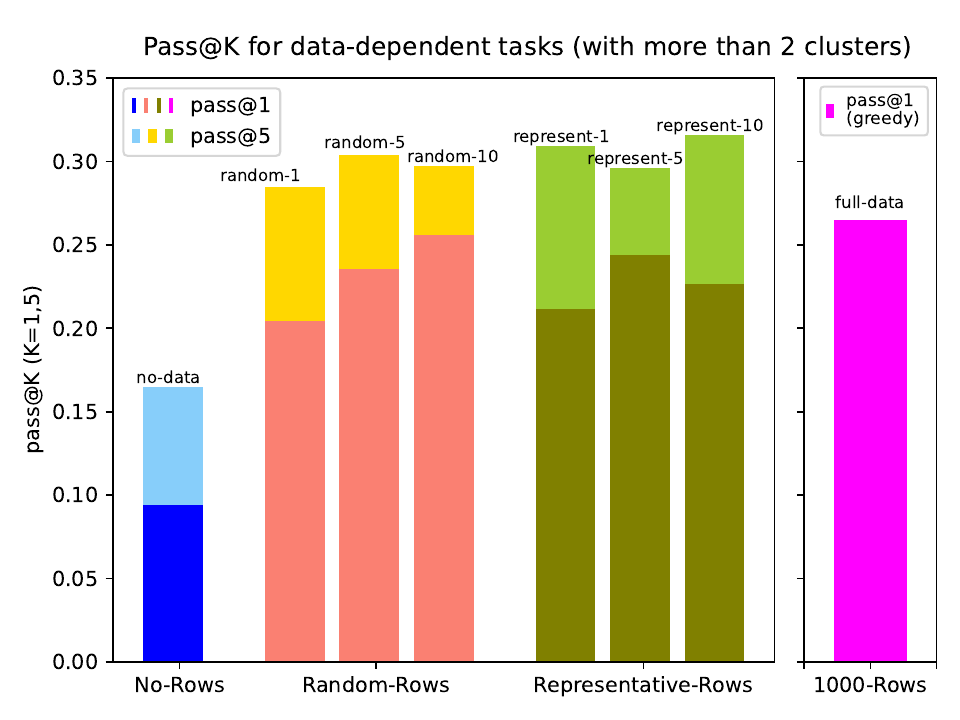} 
\caption{\pass{k} for 39\% (17/44) \dep tasks (with more than two clusters) with no-data, random selection (random-n), representative selection (represent-n) and pass@1 with greedy sampling for full-data (1000 rows).}
\label{fig:subset-clusters}
\end{figure}

\myparabf{Does model performance vary with the amount of data passed for different task classes?}
\autoref{fig:sof-data} shows the impact of the amount of data on LLM performance,
first for the entire dataset and then split by task classes. 
We see a larger drop in performance with reduced (and no) data on \dep (and \ext) tasks compared to \ind tasks.
%
Specifically, the performance gap (pass@5) between first-row and  no-data regimes is larger for the \dep and \ext classes (33.8\% and 83.5\% resp) compared to only 7.1\% for \ind tasks. 
The fact that there is any performance drop for \ind tasks indicates that having data helps the model even when the problem can be solved independently of data.
In the absence of data, almost no \ext task is solved (pass@1) but performance improves when a single row is passed.

\myparabf{Is our cluster-then-select technique effective on larger input tables?}
We evaluate our \emph{cluster-then-select} technique on Kaggle-augmented \dep tasks (with 1000 rows) since we expect to see the benefit of our approach more clearly on tasks dependent on data.
In order to do so, we compare our representative selection strategy against \emph{random} selection where the rows are randomly selected from the input table.
Among \dep tasks, we further focus on 17 (out of 44) that have input columns with at least three clusters, since with two clusters or fewer we do not expect to see much difference between the representative and random samples.
We also evaluate against two baselines: no-data (0 rows) and full-data (1000 rows). 
We run the random selection experiments five times.

\begin{figure}[t]
\centering
\includegraphics[width=0.5\textwidth, height=0.42\textwidth]{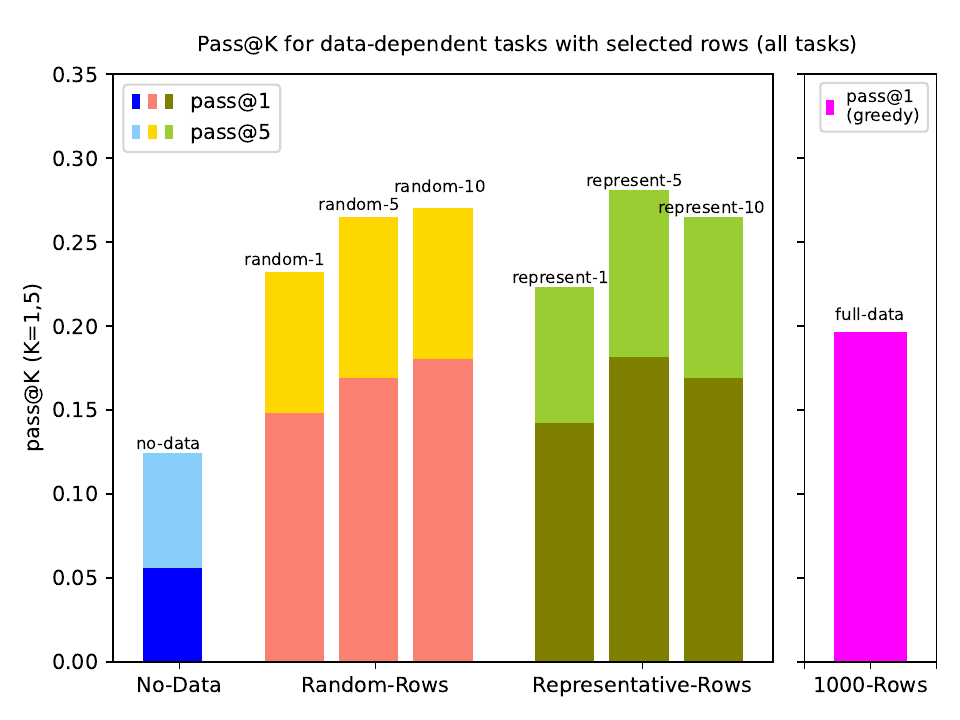} 
\caption{\pass{k} for all \dep tasks with no-data, and n=1, 5 and 10 rows passed to the model, using random (random-n), representative selection (represent-n). The completions are evaluated on 1000 rows.}
\label{fig:all-clusters}
\end{figure}
\autoref{fig:subset-clusters} shows that the model performs best with 10 most representative rows added to the prompt (pass@5 = 0.32 for represent-10).
Representative selection performs better than random selection for the same number of rows.
Specifically, represent-1 and represent-10 outperform random-1 and random-10 by 8\% and 6\% resp.
In addition, random selection has \emph{high variance},
especially for a small number of rows
(\eg \pass{1} for random-1 varies from 0.20 to 0.31 across the five runs),
which is not surprising, since the random strategy might select rows from different clusters or from the same one.
Thus, while random selection gives comparable results on average, our cluster-then-select technique offers a more consistent approach to provide the model a representative sample of the data.
Further, the low pass@k for the no-data baseline suggests that our dataset was not part of the training data, as then the model would likely perform well even without data input.
We note that while we evaluate on 1000 rows, the same cluster-then-select technique could easily scale to datasets with over 100K rows without much overhead.
We also present the evaluation results on all the 44 \dep tasks in \autoref{fig:all-clusters}.
We see that represent-5 has the highest \pass{k} for both k = 1 and 5.
Since these results include problems with fewer than three clusters, selection of even 5 representative rows boosts performance.
Notably, represent-5 also outperforms random-10.

\myparabf{Does the position of data rows in the prompt also affect performance?}
%
For the full-data baseline, we used a longer-context version of \gptfour (32k) with temperature 0 (\emph{greedy selection to eliminate variance in generations}) for the \dep tasks.
The right side of \autoref{fig:subset-clusters} shows \pass{1} for this setting with ten runs: we permute the 1000 rows in the dataframe ten times, in order to measure the sensitivity of the model to row positioning.
We observe a high variance in \pass{1} values, ranging from 0.20 to 0.32 with an average of 0.26. 
This shows that the position of rows in the dataframe influences completion quality, which aligns with previous findings about positional biases in prompts~\cite{liu2023lost}.
Surprisingly, the full-data setting (irrespective of row ordering) performs worse than selecting one random row \emph{in some cases} (\pass{1} for one random row ranges from 0.12 to 0.27 with an average of 0.20).
Note that we only report \pass{1} results for the full-data (1000 rows) setting.\footnote{\codellama results are \autoref{fig:llama-baseline}, \autoref{fig:llama-all-clusters}, \autoref{fig:llama-subset-clusters}.}%

\section{Conclusion and Future Work}
Our work highlights the importance of data for code generation on data-centric tasks and proposes a new dataset for evaluation of data-centric tasks. 
We show that providing even one data row to the model boosts performance compared to a no-data baseline.
Since providing the entire input table is often infeasible, we propose a \emph{cluster-then-select} prompting technique that selects representative rows from the data to be added to the prompt.
While randomly selecting rows also performs well, for data with a high degree of syntactic variation, it is more beneficial to add representative rows to the prompt.
For future work, handling a broader problem space (\eg, multi-table inputs, hierarchical table inputs) raises interesting challenges.
\section{Limitations} 
We discuss the limitations of our work in terms of the \sof dataset, the \emph{cluster-then-select} technique and the models used for evaluation.
Although starting from actual user-specified problems gives our results greater alignment with real spreadsheet user problems, the form that such queries take pose some potential limitations to our analysis. 
Users usually only show relevant columns of data in their queries when in actuality there might be many more unrelated columns in real spreadsheets. 
We have seen promising results applying LLMs to data tables with columns that are extraneous to the query but we do not perform a rigorous evaluation of the same. 
Furthermore, since we have collected only English queries from StackOverflow, our results may not generalize to other languages.

Our cluster-then-select prompting technique is based on the regular expression synthesis algorithm from \cite{padhi2018flashprofile}.
Given that the clusters for the input data columns are defined by the specificity of this regex synthesis, using a different clustering algorithm could potentially result in a different set of clusters.
Finally, since we draw our conclusions from the generations produced by \gptfour, future models might invalidate our conclusions.
Furthermore access to models such as \gptfour cannot be taken for granted and the costs of running our evaluation are considerable. 
Even open source models like \codellama require GPU resources for evaluation.
\section{Broader Research Impact}
To the best of our knowledge, research on prompting large language models to solve data-centric tasks with tabular data is infrequent, despite the considerable importance of such scenarios. 
Solving the problem of how to help LLM reason over large amounts of data is essential to the future of assisted decision making. 
Generating multi-step programs that require reasoning is the beginning of this journey and to make progress the community needs challenging real-world datasets to evaluate on.
By releasing our new dataset, sharing the analysis results of our experiments and releasing our prototype tool\footnote{Details discussed in \autoref{appendix_c} and \autoref{appendix_d}.}, we offer valuable benchmarks and a baseline to the wider research community which promises to encourage further exploration.
\section{Ethics Statement}
\label{sec:ethics_statement}

There are broad ethical impacts resulting from the creation of AI models that attempt to generate code solutions from natural language descriptions and these are discussed in detail in previous papers including \codex~\cite{https://doi.org/10.48550/arxiv.2107.03374}, \alphacode~\cite{li2022competition}, and \palm~\cite{PaLM}. 
These impacts include over-reliance, misalignment between what the user expressed and what they intended, potential for bias and under/over representation in the model results, economic impacts, the potential for privacy and security risks, and even environmental considerations. 
All of these considerations also apply to the work in this paper. Our focus is to highlight how the presence of data improves the performance of these models but it is important to note that the quality of the data used in the prompt will impact whether the resulting generation exhibits bias, exposes private data, etc. We explore the overall impact of providing data as part of the prompt but do not conduct a more focused analysis of determining how bias in the prompt data might influence the resulting code generation, a task we leave for future work.

There is the question of the sources of data and of consent to use the data in the manner exhibited in this paper. We have reviewed each of the datasets we have included in this paper to ensure that our use is compatible with the intent of the authors and publishers. Our datasets have also been reviewed by our institution's ethics board to review that this is an ethical use.

This paper does not directly contribute to a tool built on the assumed capabilities of language models to understand data, but nonetheless, it is motivated by their potential applications in such tools. 
These tools may be deployed in many data applications such as databases, spreadsheets, and business intelligence applications. Depending on the audience of the tool, various interaction design concerns arise. 
Explainability of the model is a key consideration, and the tool should offer decision support to evaluate mispredictions and potential next steps \cite{sarkar2022explainable}. 
Previous research of non-experts using inference driven tools for data manipulation has shown the importance of tool design in the critical appreciation of the model and its limitations, and in the potential cost of errors \cite{williams2020understanding,sarkar2015interactive}. 
As an exploratory paper without a concrete application, we do not encounter these issues, but the project has nonetheless been reviewed by our institution's ethics board.
\appendix
\bibliography{custom}
\section{Community Data License Agreement - Permissive - Version 2.0} 
This is the Community Data License Agreement - Permissive, Version 2.0 (the "agreement"). Data Provider(s) and Data Recipient(s) agree as follows:

\subsection{Provision of the Data}

\begin{itemize}
    \item A Data Recipient may use, modify, and share the Data made available by Data Provider(s) under this agreement if that Data Recipient follows the terms of this agreement.
    \item This agreement does not impose any restriction on a Data Recipient's use, modification, or sharing of any portions of the Data that are in the public domain or that may be used, modified, or shared under any other legal exception or limitation.

\end{itemize}

\subsection{Conditions for Sharing Data}

\begin{itemize}
    \item A Data Recipient may share Data, with or without modifications, so long as the Data Recipient makes available the text of this agreement with the shared Data.
\end{itemize}

\subsection{No Restrictions on Results}

\begin{itemize}
    \item This agreement does not impose any restriction or obligations with respect to the use, modification, or sharing of Results.
\end{itemize}

\subsection{No Warranty; Limitation of Liability}

\begin{itemize}
    \item All Data Recipients receive the Data subject to the following terms:
\end{itemize}
THE DATA IS PROVIDED ON AN "AS IS" BASIS, WITHOUT REPRESENTATIONS, WARRANTIES OR CONDITIONS OF ANY KIND, EITHER EXPRESS OR IMPLIED INCLUDING, WITHOUT LIMITATION, ANY WARRANTIES OR CONDITIONS OF TITLE, NON-INFRINGEMENT, MERCHANTABILITY OR FITNESS FOR A PARTICULAR PURPOSE.

NO DATA PROVIDER SHALL HAVE ANY LIABILITY FOR ANY DIRECT, INDIRECT, INCIDENTAL, SPECIAL, EXEMPLARY, OR CONSEQUENTIAL DAMAGES (INCLUDING WITHOUT LIMITATION LOST PROFITS), HOWEVER CAUSED AND ON ANY THEORY OF LIABILITY, WHETHER IN CONTRACT, STRICT LIABILITY, OR TORT (INCLUDING NEGLIGENCE OR OTHERWISE) ARISING IN ANY WAY OUT OF THE DATA OR RESULTS, EVEN IF ADVISED OF THE POSSIBILITY OF SUCH DAMAGES.

\subsection{Definitions}

\begin{itemize}
    \item "Data" means the material received by a Data Recipient under this agreement.
    \item "Data Provider" means any person who is the source of Data provided under this agreement and in reliance on a Data Recipient's agreement to its terms.
    \item "Data Recipient" means any person who receives Data directly or indirectly from a Data Provider and agrees to the terms of this agreement.
    \item "Results" means any outcome obtained by computational analysis of Data, including for example machine learning models and models' insights.
\end{itemize}
\section{Software License Agreement}
MIT License

All rights reserved. Permission is hereby granted, free of charge, to any person obtaining a copy
of this software and associated documentation files (the "Software"), to deal
in the Software without restriction, including without limitation the rights
to use, copy, modify, merge, publish, distribute, sublicense, and/or sell
copies of the Software, and to permit persons to whom the Software is
furnished to do so, subject to the following conditions:

The above copyright notice and this permission notice shall be included in all
copies or substantial portions of the Software.

THE SOFTWARE IS PROVIDED "AS IS", WITHOUT WARRANTY OF ANY KIND, EXPRESS OR
IMPLIED, INCLUDING BUT NOT LIMITED TO THE WARRANTIES OF MERCHANTABILITY,
FITNESS FOR A PARTICULAR PURPOSE AND NONINFRINGEMENT. IN NO EVENT SHALL THE
AUTHORS OR COPYRIGHT HOLDERS BE LIABLE FOR ANY CLAIM, DAMAGES OR OTHER
LIABILITY, WHETHER IN AN ACTION OF CONTRACT, TORT OR OTHERWISE, ARISING FROM,
OUT OF OR IN CONNECTION WITH THE SOFTWARE OR THE USE OR OTHER DEALINGS IN THE
SOFTWARE.

\section{Our Prototype Tool}\label{appendix_c}
\begin{figure*}[t]
    \centering
    \footnotesize
\includegraphics[width=\textwidth, height=0.52\textwidth]{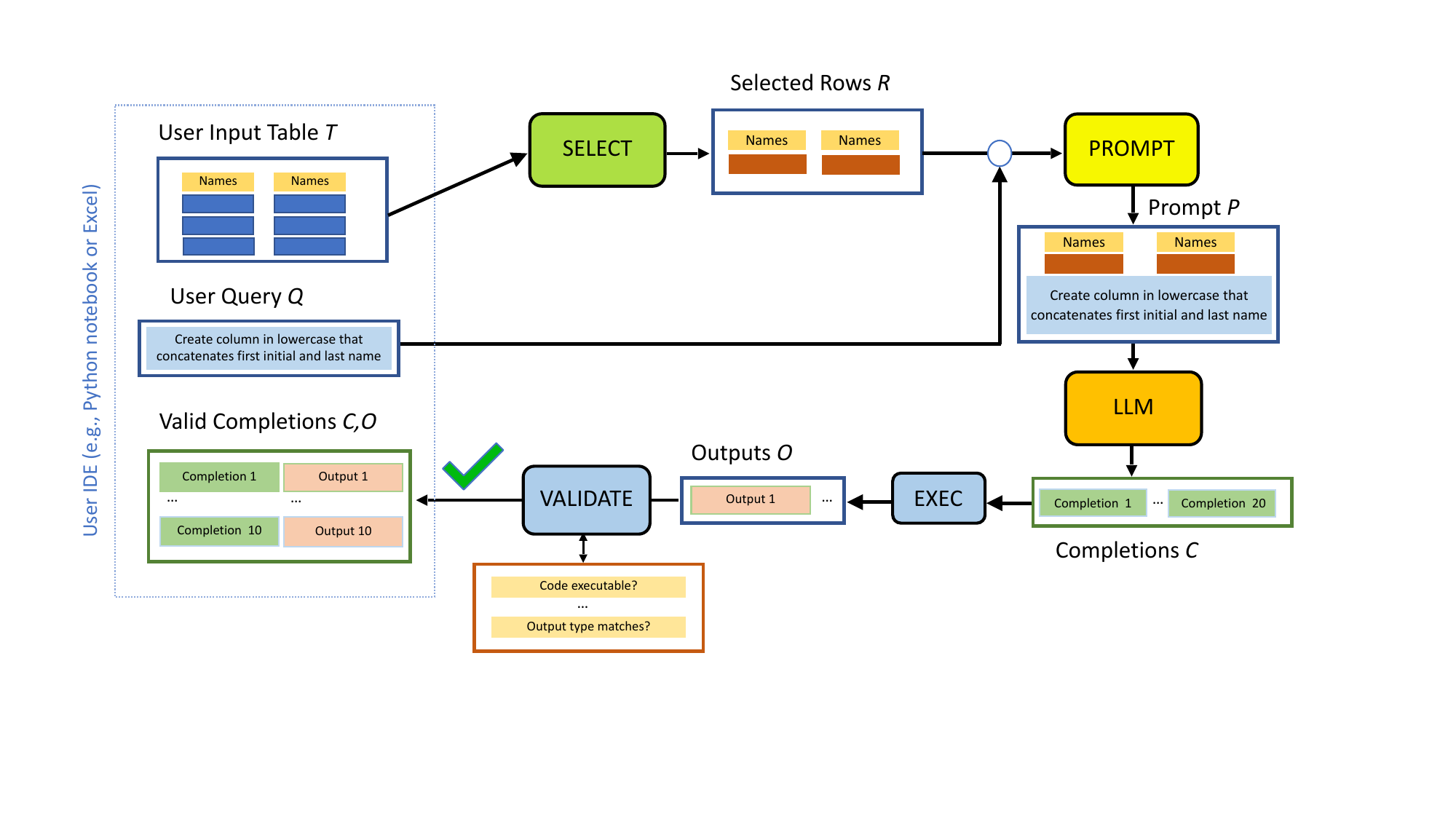}
    \vspace{-60pt}
	\caption{Our tool transforms an input table and a query into a list of valid completions. The input data is used to extract the selected rows $R$. The resulting rows and query are used to construct a prompt which is fed to a code synthesis LLM, such as \gptfour or \codellama, generating multiple possible completions. The outputs of these completions are then validated and the first $\kvar$ valid completions (along with the outputs) are returned.}
	\label{fig:codexdata}
\end{figure*}

\begin{algorithm}[h]
\footnotesize
\caption{Inference Algorithm}\label{algo-main}
\begin{algorithmic}[1]
\Require{Explicit: query $\query$, input table $\dataframe$, cardinality $\kvar$.
Implicit: completion limit $\maxcomp$ (with $\kvar \leq \maxcomp)$, number $n$ of rows to be selected.}
\Ensure{Pair of lists $(\solset,\cache)$, with $|\solset| = |\cache| \leq \kvar$, of unique completions and their corresponding outputs.}
\Procedure{Infer}{$Q,\dataframe,\kvar$}
\State $\clustermap \gets \cluster(\dataframe)$ \Comment{cluster input rows}
\State $\rows \gets \select(\dataframe, \num, \clustermap)$\Comment{select $n$ representative rows}
\State $\prompt \gets \textproc{Prompt}($\query$, \rows)$ \Comment{prompt creation}
\State $\budget, \solset, \cache \gets \maxcomp, [], []$ \Comment{initialize budget, caches}
	\While{$\budget>0 \wedge |\solset|<\kvar$ }
 		\State $\sol \gets \textproc{\llm}(\prompt)$ \Comment{sample completion}
		\State $\budget \gets \budget - 1$ \Comment{decrement budget}
\State $o \gets \textproc{Exec}(\sol, \dataframe)$  \Comment{execute against table $\dataframe$}
\If{$ \valset(o) \land (\sol \notin \solset)$}
\State $\solset \gets \solset + [\sol]$ \Comment{append completion to $\solset$}
\State $\cache \gets \cache + [o]$ \Comment{append output to $\cache$}
\EndIf
\EndWhile
\State \textbf{return} $(\solset,\cache)$
\EndProcedure
\end{algorithmic}
\end{algorithm}	

The high-level workflow of our tool is depicted in ~\autoref{fig:codexdata} and formalized in Algorithm~\ref{algo-main}.
The tool takes as input a query $\query$ expressed in natural language, an input table $\dataframe$ as a Pandas dataframe, and the target cardinality $\kvar$ of distinct completions to generate. 
We set a limit $\maxcomp$ on the number of calls to $\llm$  ($\maxcomp = 8 \kvar$).
For our running example, $\kvar$ is 1,
$\query$ is ``create a new column in lowercase that concatenates the first initial and the last name.'',  
and $\dataframe$ is 
\scode{Data(\{"\tcode{Names}":["John Smith", "Jack Will Anders", ...]\})}.
At a high-level, the algorithm first clusters the data in $\dataframe$ based on automatically synthesized regular expressions and stores them in a map $\clustermap$ (line 2). 
It then extracts \emph{representative rows} of the table using \textproc{Select} (line 3); combines the query $\query$ and the rows $\rows$ to create a prompt $\prompt$ using~\textproc{prompt} (line 4); and then queries $\llm$ repeatedly using this prompt until the target completions are reached or we exceed the budget of calls (lines 7-12).
Each completion $c$ is executed on the input table (line 9) using an \textproc{exec} procedure, and if the completion is new and its output $o$ satisfies a \textproc{Validate} procedure, the two are accumulated in $C$ and $O$ which are then returned.
We describe each of the procedures in detail below.

\setulcolor{green}
\myparabf{\ul{$\cluster$}}
This procedure clusters the rows in the input table $\dataframe$ based on their syntactic structure.
To capture the syntactic variation among input rows, we rely on an existing tool ~\cite{padhi2018flashprofile}, which takes as input a set of strings and synthesizes a set of regular expressions (regexes) from a restricted class, such that each input string matches one of the regexes.
In our example, the tool synthesizes four regexes:
\scode{[A-Z][a-z]+[$\textbackslash\textbackslash$s] [A-Z][a-z]+}, for rows with no middle name like "John Smith",
and similar regexes for rows with dashed last names like "Ashley Kelsey-Poe", and one or more middle names.
These regexes are then used to cluster input strings.

\setulcolor{green}
\myparabf{\ul{$\select$}}
The \textproc{Select} procedure selects the top-n most representative rows from the input table.
We frame the selection of most representative rows as a \emph{weighted maximal coverage problem}--- a well-known NP-complete problem~\cite{maxcoverage} that can be solved approximately using the greedy algorithm in Algorithm~\ref{alg:select}. 
The algorithm takes as input the table $\dataframe$, a map $\clustermap$ from the rows of the table to the set of clusters covered by the element in each column of the row. 
It also takes as input the row budget $n$. 
The algorithm iterates over all rows in $\dataframe$ not already in $\rows$ (line 3) and in each iteration selects the row whose elements maximize the size of clusters covered (line 4), adding this row to $\rows$.

\begin{algorithm}[t]
\footnotesize

\caption{Rows Coverage Algorithm $\select$}
\label{fig:algo-select}	
	\begin{algorithmic}[1]
\Procedure{Select}{$\dataframe$, $\num$, $\clustermap$}
\While{$\left | \rows  \right |$ < $n$}
\For{$r \in \dataframe_{r} \wedge r \notin \rows$} \State$\textsc{best}\gets\textbf{argmax}(\sum \set{\left|c_{i}\right|}{c_{i}\in\clustermap[r]})$

\Comment{greedily increase coverage}

\EndFor
\State $\rows \gets \rows \cup \textsc{best}$

\EndWhile 
 
\Return $\rows$
\EndProcedure
\end{algorithmic}\label{alg:select}
\end{algorithm}	

\setulcolor{yellow}
\myparabf{\ul{$\textproc{Prompt}$}} 
The prompt creation procedure~$\textproc{Prompt}$ creates a textual prompt by concatenating the NL query and the representative rows $\rows$ which are in form of a Pandas dataframe.
An example prompt is in Appendix~\ref{prompt-template}.

\setulcolor{orange}
\myparabf{\ul{$\textproc{\llm}$}}
The completion procedure $\textproc{\llm}$ queries \gptfour (or another code-generating model), passing the prompt $\prompt$ and also the predefined stop sequences.
We use stop sequences that we have found to allow the LLM to generate at least one solution while typically not using the entire token budget. 
Note that the LLM needs to produce multiple completions, because it will filter out invalid completions.
A naive approach would be to request a single completion, validate it,
and repeat the process until $k$ distinct valid completions are obtained;
this, however, requires sending the prompt to the LLM every time, 
which incurs a monetary cost.
An alternative approach is to \emph{batch} the completions, 
\ie request some number $b$ of completions in parallel;
if the batch size $b$ is too large, however,
this also incurs unnecessary cost,
since we are requesting more output tokens than we need.
Details in Appendix~\ref{valid-completions}.

\setulcolor{cyan}
\myparabf{\ul{$\textproc{Exec}$}}
The procedure $\textproc{Exec}$ turns each LLM completion 
into a stand-alone executable program
and runs it to obtain the final output $o$.
There are two main challenges to be addressed in this step.
First, LLM completions do not have a consistent way of identifying the final output:
for example, 
the last line of the completion might be an expression that computes the output,
or an assignment to a \scode{result} variable,
or a print statement.
So our tool uses a predefined set of rewrite rules,
which we developed by analyzing the patterns in completions.
The second challenge is that executing arbitrary LLM-generated code poses a security risk;
for this reason, we execute completions in a sandbox.
Further details are available in Appendix~\ref{exec-completions}.

\myparabf{\ul{$\textproc{Validate}$}}
The procedure $\valset$ checks that the output value $o$
is a dataframe with the right dimensions.
The completions that executed without runtime errors during $\textproc{Exec}$
and passed the output validation
are deemed \emph{valid}.
Further details are available in Appendix~\ref{validate-completions}.
\section{Experimental Details} \label{appendix_d}
\subsection{\codellama Results}
\begin{figure}[t]
\centering
\includegraphics[width=0.45\textwidth, height=0.38\textwidth]{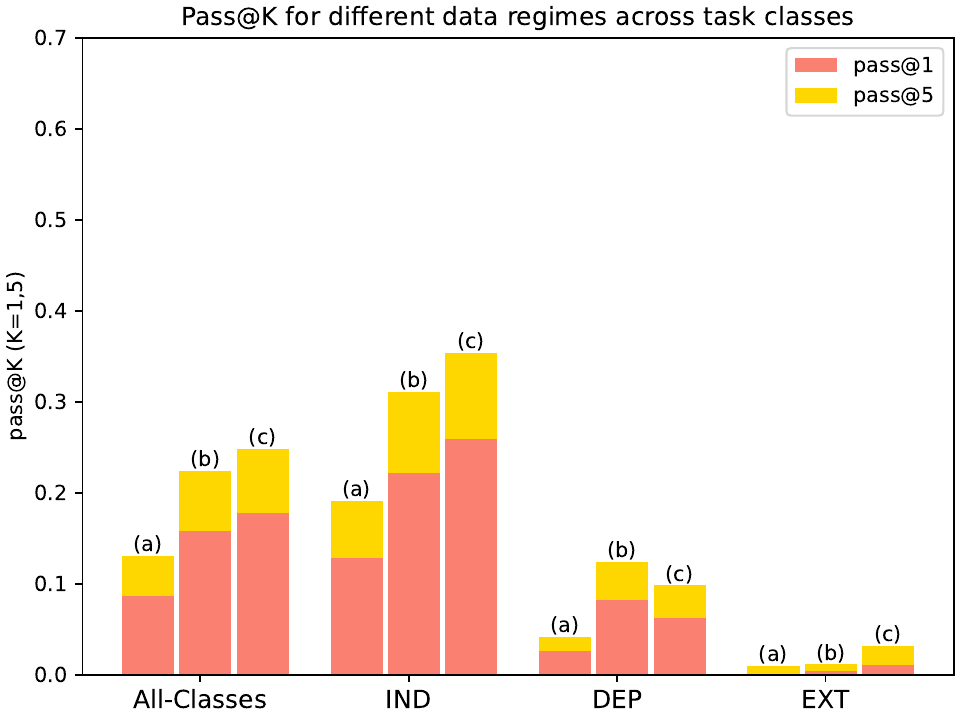} 
 \caption{\pass{k} (for \codellama) with (a) no-data, (b) first-row, and (c) full-data (10 rows) passed to the model. The leftmost group of bars represent pass@k with all classes followed by separate pass@k for \ind, \dep and \ext tasks. Smaller models have a huge performance drop. But the trend of performance improving with the amount of data passed to the model is seen.}
\label{fig:llama-baseline}
\end{figure}

\begin{figure}[t]
    \centering
    \includegraphics[width=0.48\textwidth, height=0.38\textwidth]{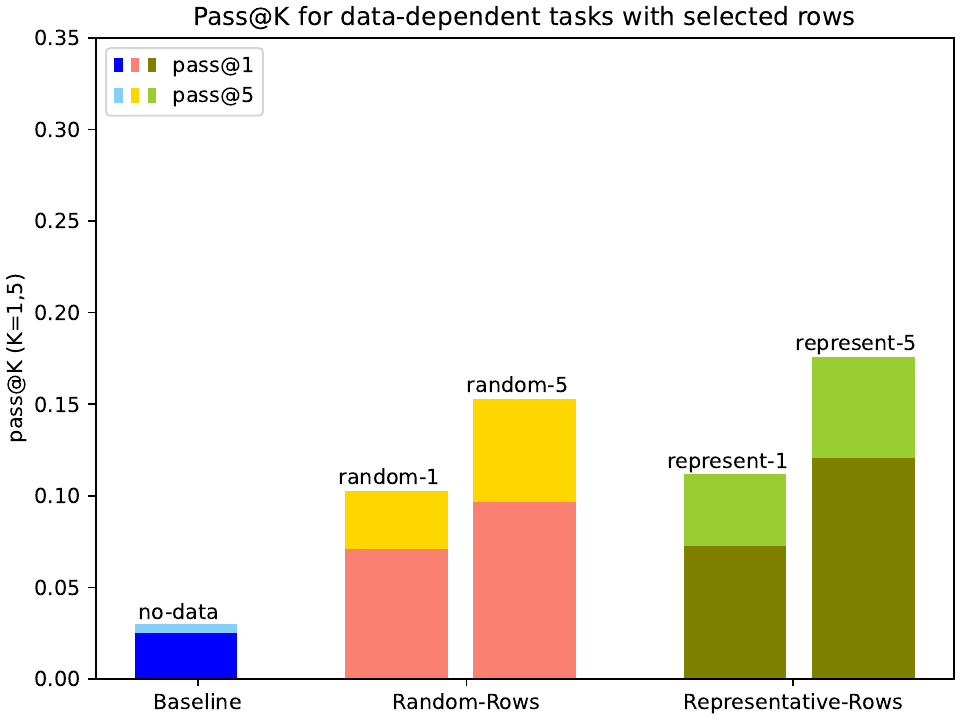}
    \caption{\pass{k} (for \codellama) for all 44 \dep tasks with no-data, and n=1 and 5 rows passed to the model, using random (random-n) selection, representative selection (represent-n) and full-data (1000 rows). Completions are evaluated on 1000 rows.}
    \label{fig:llama-all-clusters}
\end{figure}

\begin{figure}[t]
    \centering
    \includegraphics[width=0.48\textwidth, height=0.38\textwidth]{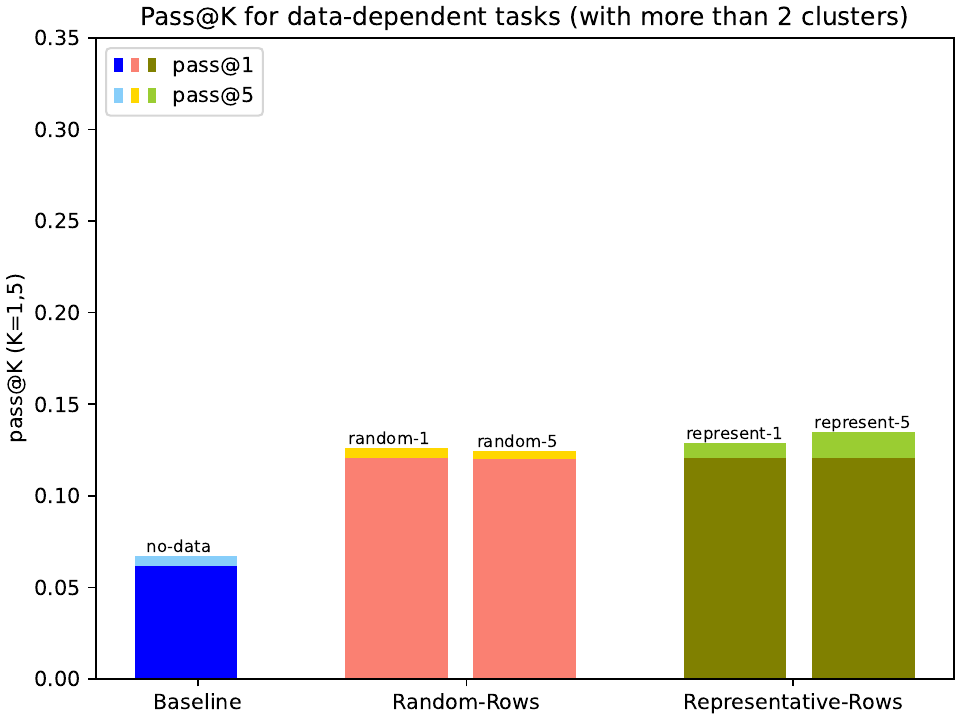}
    \caption{\pass{k} (for \codellama) for 17 out of 44 \dep tasks (more than two clusters) with no-data, random selection (random-n) and representative selection (represent-n). Completions are evaluated on 1000 rows.}
    \label{fig:llama-subset-clusters}
\end{figure}

We do a performance comparison for no-data, first-row and full-data regimes and the different selection strategies with \codellama \cite{roziere2023code} as the LLM.
The results with \codellama are presented in \autoref{fig:llama-baseline}, \autoref{fig:llama-all-clusters} and \autoref{fig:llama-subset-clusters}.
\subsection{Evaluation Metrics}
The probability that at least one of $k$ inferred outputs is correct is called \pass{k} \cite{https://doi.org/10.48550/arxiv.2107.03374}.
More formally, pass@$k$ is the probability that with a sample of $k$ code completions, at least one is correct.
To measure this probability empirically for each datapoint, we compute up to $m$ valid programs by sampling from the LLM (\gptfour or \codellama).
We count the number $s$ of correct completions, and hence compute an estimate of \pass{k} as $1-\binom{m-s}{k}/\binom{m}{k}$ \cite{https://doi.org/10.48550/arxiv.2107.03374}.
By computing $m > k$ completions the estimate has lower variance than by simply computing $k$ completions.
Each \pass{k} on a whole dataset is the average of \pass{k} over all its datapoints. 
All evaluation results are averaged over tasks, computing $m$ valid completions to estimate \pass{k} or \partialpass{k}{X}. 
In practice, we set $m=20*k$ when we report results for $k=1$ or $k=5$. 

\subsection{Prompt Template} \label{prompt-template}

For each task, we generate prompts according to the data regimes and selection strategies as described above. An example prompt for the query "Create a new column with the difference 
in hours, minutes and seconds between the 
two timestamps in the format HH:MM:SS" with one row selected:
\definecolor{codegreen}{rgb}{0,0.6,0}
\definecolor{codegray}{rgb}{0.5,0.5,0.5}
\definecolor{backcolour}{rgb}{0.95,0.95,0.92}

\lstdefinestyle{mystyle}{
    backgroundcolor=\color{backcolour},   
    commentstyle=\color{codegreen},
    numberstyle=\tiny\color{codegray},
    basicstyle=\ttfamily\footnotesize,
    breakatwhitespace=false,         
    breaklines=true,                 
    captionpos=b,                    
    keepspaces=false,                 
    numbers=left,                    
    numbersep=5pt,                  
    showspaces=false,                
    showstringspaces=false,
    showtabs=false,                  
    tabsize=2
}
\lstset{style=mystyle}

\begin{lstlisting}[language=Python, caption=Example of a prompt]
import pandas as pd
df = pd.DataFrame()
df[`Start'] = [`2/22/2015 1:06:20 PM']
df[`End'] = [`2/23/2015 3:08:20 PM']
#Create a new column with the difference in hours, minutes, and seconds between the two timestamps in the format HH:MM:SS
\end{lstlisting}

\subsection{Generation of Completions} \label{valid-completions}

\mypara{Parallelization}~For efficiency, we request multiple completions from \gptfour per iteration.
To try to minimize both inference time and the load on OpenAI's servers, we adapt the batch size to an estimate of the probability that the next completion is valid. 
The batch size used in each iteration is $n = \min\left(\lceil r/p \rceil, \budget, L\right)$, where $r = \kvar-|\solset|$ is the number of valid completions still to obtain, $\budget$ is the remaining completion budget, and $L$ is a parallelization limit enforced by the \gptfour API. 
The probability estimate $p$ is updated after each iteration by counting the number of valid and invalid completions in that iteration's batch.
Since \pass{k} is calculated only from valid completions, it is not influenced by either parallelization or batch size adaptation. 

\mypara{Stop sequences}~The most effective stop sequence we found that allows \gptfour to generate at least one solution while not usually using the entire token budget is a blank line followed by a line comment; i.e.\ \scode{\textbackslash n\#}. 
Further, to keep \gptfour from generating what appears to be the rest of a forum post after a code snippet, we also use the stop sequence \scode{</code>}.

\mypara{Completion cleanup}~Since \gptfour's training data likely contains forum posts, some completions would raise \scode{SyntaxError} exceptions when executed due to formatting artifacts, and therefore be invalid.
Instead, to make the most of the completion budget, we replace formatting artifacts \ie we replace HTML escape sequences such as \scode{\&lt;} and \scode{\&quot;} with Python operators and delimiters.
Cleanup also removes unnecessary whitespace, blank lines, comments, and truncates completions at \scode{\textbackslash n\#} when it appears after executable code.

\subsection{Execution of Completions} \label{exec-completions}

\mypara{Rewriting}~Completions returned by \gptfour do not clearly indicate which variables or expressions are intended to be the answer to a query. 
This must be inferred from the shape of the code. We found that an effective way to identify and expose the likely answer is to search backwards to find the last unindented (i.e.\ top-level) statement that has one of a few forms, and rewrite the completion so that its last statement is an assignment to a fresh identifier \scode{var\_{out}}. 
The statement forms and rewrites are
\begin{itemize}
    \item \scode{var = expr}: append the statement \scode{var\_{out} = var} to the completion.
    \item \scode{var[expr\_i] = expr}: append the statement \scode{var\_{out} = var} to the completion
    \item \scode{print(expr, ...)}: replace this statement and the rest of the completion with \scode{var\_{out} = expr}
    \item \scode{expr}: replace this statement and the rest of the completion with \scode{var\_{out} = expr}
\end{itemize}

Rewriting also inserts \scode{import} statements for common libraries (e.g.\ \scode{import numpy as np}).
The rewritten completion is appended to the code that defines the input dataframe to create a complete program. 
The program and output variable \scode{var\_{out}} are sent to a sandbox for execution.

\mypara{Sandboxing}~Because of security risks inherent in running the LLM-generated code, we run completed programs in a sandbox. 
Our sandbox is a JavaScript web service that runs Python programs in Pyodide~\cite{michael_droettboom_2022_5834941}, a Python distribution for WebAssembly. 
While Python programs running in Pyodide have access to the host's network resources, they at least are isolated from other host resources including its filesystem, offering some level of protection from malicious or accidentally harmful completions.
After running the code, the sandbox returns the value of \scode{var\_out}.

\subsection{Validation of Completions}\label{validate-completions}

For a completion to be considered a correct solution in the calculation of \pass{k}, its actual output must match the expected output. 
Matching cannot be the same as equality and still conform to a reasonable notion of correctness; for example, the natural breakdown of a solution might generate intermediate columns in the actual output that are not in the expected output.
The actual output is allowed to vary from the expected output in the following ways and still match the expected output:
\begin{itemize}
    \item Extra columns
    \item Different column order
    \item Different column headers
    \item Number expected; actual is a number within small relative error (default 0.01)
    \item Number expected; actual is a string that parses as a number within small relative error
    \item Boolean expected; actual is number 0 or 1
    \item Boolean expected; actual is a string that represents a truth value
    \item String expected; actual is a string that differs only in case
\end{itemize}
Allowed string truth value representations, allowed relative error, and whether string matching is case-sensitive are (optionally) overridden per data point as appropriate.
\end{document}